\documentclass[a4paper]{jpconf}
\newcommand{\beq}{\begin{equation}}
\newcommand{\eeq}{\end{equation}}
\newcommand{\bear}{\begin{eqnarray}}
\newcommand{\eear}{\end{eqnarray}}
\newcommand{\mrm}[1]{\mathrm{#1}}
\newcommand{\hst}{\mathcal H}
\newcommand{\est}{\mathcal E}
\newcommand{\om}{\omega}

\renewcommand{\vec}[1]{\mathbf{#1}}

\usepackage{graphicx}

\begin{document}

\title{Superconducting cavity transducer for resonant gravitational radiation antennas}

\author{R Ballantini$^1$, M Bassan$^2$, A Chincarini$^1$, G Gemme$^1$ and R Parodi$^1$ and R Vaccarone$^1$}

\address{$^1$ INFN, Sezione di Genova, via Dodecaneso, 33, I-16146, Genova, Italy}
\address{$^2$ Universit\`a degli studi di Roma "Tor Vergata", via della Ricerca Scientifica, 1, I-00133, Roma, Italy}

\ead{gianluca.gemme@ge.infn.it}

\begin{abstract}
Parametric transducers, such as superconducting rf cavities, can boost the bandwidth and sensitivity of the next generation resonant antennas, thanks to a readily available technology. We have developed a fully coupled dynamic model of the system "antenna--transducer" and worked out some estimates of signal--to--noise ratio and the stability conditions in various experimental configurations. We also show the design and the prototype of a rf cavity which, together with a suitable read--out electronic, will be used as a test bench for the parametric transducer.
\end{abstract}

\section{Introduction}
A first--generation parametric transducer was designed, constructed and extensively tested at the University of Western Australia in Perth \cite{blair_cavity, blair_antenna, blair_signal, blair_analysis1, blair_analysis2}. Another device of this kind is now being developed for the Brazilian spherical gravitational wave detector SCHENBERG \cite{Barroso:2004vz,Ribeiro:2004wa}.

This device is based on a re--entrant cavity whose resonant frequency is modulated by the gap spacing between the cavity and the antenna. The change in frequency due to the motion creates modulation sidebands in the output signal. Additional amplification of the high--frequency signal is usually necessary. 

\begin{figure}[hbt]
\begin{center}
\includegraphics[scale=0.5]{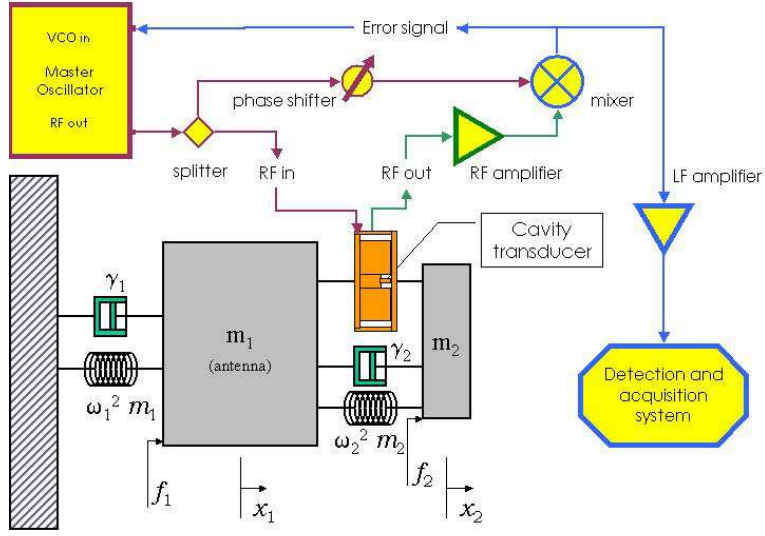}
\caption{
\label{fig:modello}
Detector conceptual layout}
\end{center}
\end{figure}

The resonant antenna (mass $m_1$) and the secondary mass $m_2$ (with $m_2\ll m_1$) form a system of two coupled harmonic oscillators (see fig. \ref{fig:modello}). The oscillators are designed to have (when \emph{uncoupled}) almost equal resonant frequencies $\omega_1\simeq \omega_2$. Mechanical dissipation is included in the model through a dissipation constant $\gamma_i$ ($i=1,2$) from which the energy decay time constant $\tau_i=m_i/\gamma_i$, and mechanical quality factor $Q_i=\omega_i\tau_i$ can be derived.

\section{Equations of motion}
The equations that describe the dynamics of the mechanical system alone can be derived from the Lagrangian function:
\beq
\label{eq:lag_mecc}
{\mathcal L}_m =\frac{1}{2} m_1 \dot x_1^2 + \frac{1}{2} m_2 \dot x_2^2 - \frac{1}{2} m_1 \omega_1^2 x_1^2 - \frac{1}{2} m_2 \omega_2^2 \left( x_2-x_1\right)^2
\eeq
together with a Dissipative function \cite{landau_stat}: 
\beq
\label{eq:diss_mecc}
{\mathcal D}_m = \frac{1}{2} \frac{m_1 \dot x_1^2}{\tau_1} + \frac{1}{2} \frac{m_2 \left( \dot x_2-\dot x_1\right)^2}{\tau_2}
\eeq

To include in the model the dynamic interaction between the mechanical resonators and the transducer, while still keeping the equations as simple as possible, we shall make use of the fact that any configuration of the field inside the resonator can be expressed as the superposition of the electromagnetic normal modes \cite{slater}: $\vec E(\vec r,t)=(\epsilon_0)^{-1/2}\sum {\mathcal E}_n(t)\, \vec E_n(\vec r)$, $\vec H(\vec r,t)=(\mu_0)^{-1/2}\sum {\mathcal H}_n(t)\, \vec H_n(\vec r)$, 
with the time--dependendent amplitudes\footnote{The choice of the normalization constants is such that $\hst(t)$ and $\est(t)$ have the same dimension. The total energy of the field is given by $U=1/2(\hst^2+\est^2)$.} ${\mathcal E}_n(t)\equiv (\epsilon_0)^{1/2}\int\vec E \cdot \vec E_n\,\mrm{d}V$,
${\mathcal H}_n(t)\equiv (\mu_0)^{1/2}\int\vec H \cdot \vec H_n\,\mrm{d}V$, and the orthonormality conditions
$\int\vec E_n \cdot \vec E_m\,\mrm{d}V = \int\vec H_n \cdot \vec H_m\,\mrm{d}V = \delta_{nm}$.

We shall consider an experimental situation in which an external source at angular frequency $\omega$ excites the field in the cavity near one of its eigenfrequencies $\omega_0$. In this case the field stored in the cavity essentially coincides with the eigenmode at frequency $\omega_0$ and we shall write: $\vec E(\vec r,t)\simeq(\epsilon_0)^{-1/2}{\mathcal E}(t)\, \vec E_0(\vec r)$ and $\vec H(\vec r,t)\simeq(\mu_0)^{-1/2}{\mathcal H}(t)\, \vec H_0(\vec r)$.

The time--dependent amplitudes of the electromagnetic field stored in the re--entrant cavity behave like a simple harmonic oscillator with resonant frequency $\omega_0$ and energy decay time $\tau$. This harmonic oscillator can be described by e.m. terms in the Lagrangian and in the Dissipative function:
\begin{eqnarray}
\label{eq:lag_em}
{\mathcal L}_{em} & = & \frac{1}{2\omega_0^2} \,\dot \hst^2 - \frac{1}{2}\, \hst^2\\
{\mathcal D}_{em} & = & \frac{1}{2\tau\omega_0^2}\, \dot \hst^2
\end{eqnarray} 

The \emph{dynamic interaction} between the e.m. oscillator and the mechanical oscillators has to be included. In our model the transducer is coupled to the secondary mass, so the interaction term will have the form \cite{goubau,cqg}:
\beq
\label{eq:lag_int}
{\mathcal L}_{int} = -\frac{1}{2} \,\kappa\, \left ( x_2 - x_1\right ) \hst^2
\eeq
The constant $\kappa$ has dimensions $[\mrm{length}]^{-1}$.

The equations of motion of the three--coupled--oscillators system are easily derived as
\beq
\label{eq:ode1}
\ddot x_1 + \frac{\dot x_1}{\tau_1} + \omega_1^2 x_1 -\frac{m_2}{m_1}\frac{\dot\delta}{\tau_2} - \frac{m_2}{m_1}\omega_2^2 \delta - \frac{1}{2 m_1} \,\kappa\,\hst^2 = \frac{f_1}{m_1}
\eeq
\beq
\label{eq:ode2}
\ddot \delta + \frac{\dot\delta}{\tau_2} + \omega_2^2 \delta + \ddot x_1 + \frac{1}{2 m_2} \,\kappa\,\hst^2 = \frac{f_2}{m_2}
\eeq
\beq
\label{eq:ode3}
\ddot \hst + \frac{\dot \hst}{\tau}+\omega_0^2 \hst \left( 1+\kappa\,\delta\right ) = \omega_0^2 f_s
\eeq
where $\delta=x_2-x_1$.

The right hand side of the above equations accounts for possible external interactions ($f_1$ might describe the gravitational interaction coupled to the primary oscillator,  $f_2$ might describe the gravitational interaction coupled to the secondary oscillator and $f_s$ describes the external rf source at angular frequency $\omega_{rf}$). For $\kappa=0$, we recover the equations of motion of the two coupled mechanical oscillators. 

It is worth stressing again that whenever an interaction term is present in the third (e.m.) equation (the interaction term \emph{must} be present if we want to have a signal from the transducer), terms of the same order of magnitude $\sim \kappa$ are present in the mechanical equations. From this we can conclude that the transducer plays an {\em active} role in determining the dynamics of the system. At least one dynamic variable, describing the state of the transducer, has to be coupled to some degree of freedom of the monitored system (in this case the e.m. field amplitude $\hst$ is coupled to the positions $x_2-x_1$ of the vibrating masses). The same variable appears in the equations of motion of the monitored system, and its contribution cannot, in general, be neglected.

The explicit form of the coefficient $\kappa$ can be deduced from eq. (\ref{eq:ode3}). This equation describes a frequency modulated oscillator with time--varying resonant frequency $\omega^2(t)=\omega_0^2[1+\kappa\,\delta(t)]$. This term can be put in the following form:
\beq
\label{eq:slater}
\frac{\omega^2-\omega_0^2}{\omega_0^2}\simeq 2\frac{\omega-\omega_0}{\omega_0} = \kappa\,\delta
\eeq
or
\beq
\label{eq:coeff}
\kappa = \frac{2}{\omega_0}\frac{\Delta\,\om}{\Delta\, \delta}
\eeq
which shows that $\kappa$ is essentially a fractional frequency tuning coefficient which depends only on the geometry of the transducer. In the present design $\omega_0/(2\pi)\simeq 5.5\times 10^9\,\mrm{Hz}$ and $1/(2 \pi)\, \Delta \om / \Delta \delta \simeq -1.4\times 10^{14}\,\mrm{Hz/m}$, so that $\kappa\simeq 5\times 10^4$.

\section{Feedback}
The transducer is complemented with a feedback loop, which keeps the local oscillator on track with the cavity. As shown in fig. \ref{fig:modello}, the loop components are arranged in a phase comparator configuration, so that the output of the mixer is proportional to the phase difference between the rf generator signal (at angular frequency $\omega_{rf}$) and the cavity output. 

The cavity output has the same instantaneous frequency as the rf generator signal, but it is phase-shifted of an amount which, near the cavity resonance, is linearly proportional to $\omega_{rf} - \omega_c$, where $\omega_c$ is the instantaneous resonator angular frequency
\footnote{It should be remarked that $\omega_c$ is only the natural oscillation frequency of the resonator and, in general, it is different from $\omega_{rf}$, which is the frequency of the rf signal passing through the cavity. The unperturbed cavity ($\delta=0$) is designed to have the natural oscillation frequency equal to $\omega_0$. $\omega_c$ strongly depends on the geometry of the cavity and is, by design, maximally sensitive to the gap spacing.} given by 
\beq
\label{eq:feed1}
\omega_c=\omega_0\left( 1+ \frac{1}{2}\,\kappa\,\delta\right)
\eeq
For a stationary (linearized) solution, where $\delta(t)$ and $f_1(t) \propto e^{j \Omega t}$ (see eq. \ref{eq:ode1}), sidebands appear in the field solution at frequencies $\omega_{rf} + \Omega$ and $\omega_{rf} - \Omega$. 
The feedback circuit allows to keep the rf generator frequency tuned with the cavity instantaneous frequency by feeding back the error signal coming from the mixer into the Voltage Controlled Oscillator (VCO). This error signal, properly amplified, carries the useful information on the GW, already demodulated from the carrier signal. In the first order approximation, near the cavity resonant frequency, the error signal is given by
\beq
\label{eq:feed2}
V_e=-2\,G_{rf}\,K_m\,\tau\, K_c(\Omega)\frac{\beta}{(\beta+1)^2}\left( \omega_{rf}-\omega_c\right)
\eeq
where $G_{rf}$ is the gain of the rf low--noise amplifier, $K_m$ is the mixer conversion loss, $\tau$ is the energy decay time of the superconducting cavity (see eq. \ref{eq:ode3}), and $\beta$ is the cavity coupling coefficient, given by the ratio between the the cavity input impedance and the circuit impedance. $K_c(\Omega)$ describes the low--pass action of the microwave cavity on the amplitude of modulation sidebands surrounding the carrier, with cut--off frequency $1/(2 \tau) = \omega_0 / (2 Q_L)$, where $Q_L$ is the loaded electromagnetic quality factor of the cavity.
Finally the VCO instantaneous frequency is given by
\beq
\label{eq:feed3}
\omega_{rf}=\omega_f+K_V\, V_e
\eeq
where $\omega_f$ is the generator free--running angular frequency (usually chosen to be $\omega_f = \omega_0$), and $K_V$ is the VCO voltage--to--frequency conversion characteristic.
 
The feedback implementation is prone to two main noise sources, namely: the rf oscillator phase and amplitude noise and the amplifiers (rf and if) input noise. The rf amplifier is one of the most critical component in the loop, as its noise input level ultimately sets a limit on the sidebands amplitude detection. In few words, the cavity instantaneous frequency is modulated by the moving gap and therefore sidebands arise from the rf signal, with an amplitude roughly proportional to the frequency displacement causing the modulation, $\Delta \omega$, divided by the modulating frequency: $A \propto \frac{\Delta \omega}{\Omega}$. To be detected, this amplitude must be greater than the rf amplifier input noise. We should note however that state--of--the--art, commercially available amplifiers can already provide near quantum--limited, cryogenic--operated devices ($T_n \simeq 2$ K, that is $\simeq 8 \,\hbar$ @ 5 GHz).

The cavity is pumped through one port and read through the same port, that is through the reflected signal. If the electromagnetic coupling $\beta $ to the cavity resonant mode is $\beta \simeq 1$, then the reflected signal amplitude near resonance is nearly zero. In a real experimental condition, when $\beta $ can be far from 1, one might need to provide a more complicated circuitry to suppress the strong reflected signal from the cavity entering the rf amplifier (carrier suppression interferometer, see \cite{blair_signal}).

The feedback equations couple the rf oscillator frequency $\omega_{rf}$ with the masses position $\delta$ through the cavity instantaneous frequency $\omega_c$. For very high open-loop gain, the rf generator instantaneous frequency closely follows the cavity one. To extract the GW information one can use either the feedback error signal (whose typical frequencies are $\approx 10^3$ Hz) or the rf generator output ($\approx 10^9$ Hz). Both techniques are theoretically valid and the choice between them is a matter of experimental feasibility.

\begin{figure}[hbt]
\begin{center}
\includegraphics[scale=0.6]{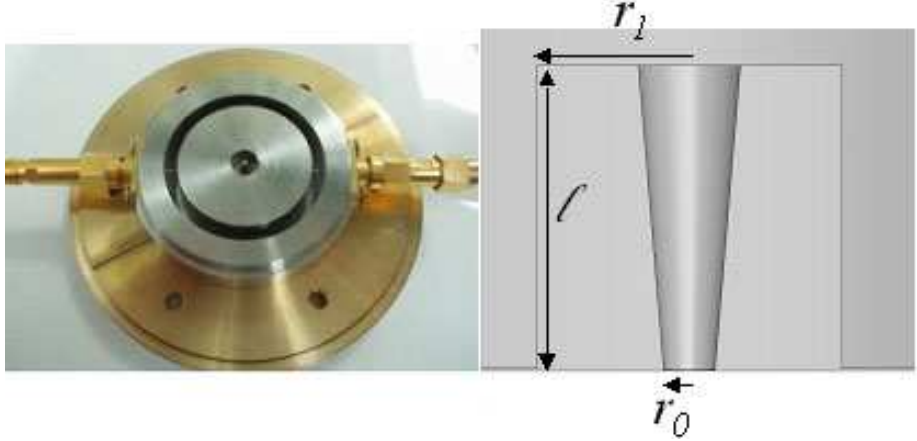}
\caption{
\label{fig:dimensioni}
The re--entrant cavity transducer and a sketch of its section. In the left picture the rf--choke designed to minimize radiation losses is visible. The choke is also used to host the input and output coupling ports. In this set--up two ports are used. In the real experimental configuration the cavity will be pumped through one port and read through the same port.}
\end{center}
\end{figure}

\section{\label{sec:cavity} The rf superconducting cavity}
The re--entrant cavity is basically a lumped elements LC resonator, where the capacitance is determined by the spacing between the central post and the end wall of the antenna, and the inductance is mainly due to the central section of the cavity. Any change in the distance between the central post and the antenna modulates the capacitance of the cavity and, as a consequence, its resonant frequency, $\om_0^2 = 1/(LC)$, producing sidebands in the pump signal which are displaced from the pump by the antenna frequency. These sidebands contain the information about the amplitude, phase and frequency of the external perturbation causing the vibration of the antenna--transducer.

In our design the cavity has a lenght $\ell = 5\,\mrm{mm}$ and radius $r_1 = 3\,\mrm{mm}$. The central post has radius $r_0 = 0.5\,\mrm{mm}$. High transducer sensitivity is obtained using a small gap $d$ and we choose $d\simeq 15 \, \mrm{\mu m}$ (see fig. \ref{fig:dimensioni}). The cavity is pumped by an external ultra--low noise rf source near its resonant frequency $f_0$.
Finite element calculations gave the following values for the cavity operating parameters: resonant frequency (TM mode) $f_0 = 5.5\,\mrm{GHz}$, sensitivity $\Delta f/\Delta \delta = 1.4\times 10^{14} \,\mrm{Hz/m}$. 

\section{Expected performance}
\begin{figure}[hbt]
\begin{center}
\includegraphics[scale=0.6]{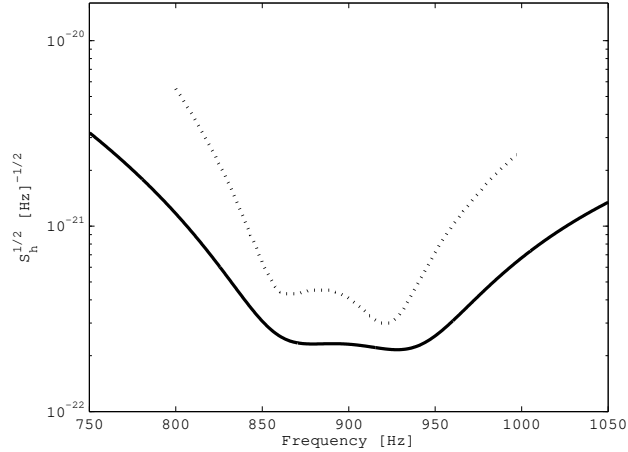}
\caption{
\label{fig:larga}
Expected performance of the parametric transducer coupled to an existing antenna (solid line), compared to the best expected performances of the SQUID readout (dotted line). The parameters used in this calculation are listed in table \ref{tab:params}. }
\end{center}
\end{figure}

The calculated performance of the parametric transducer is shown as a replacement of the single gap capacitive transducer read by a SQUID on an existing antenna. Figure \ref{fig:larga} compares the parametric readout with an advanced double SQUID readout calculated on a resonant antenna \cite{vinante}. An unloaded quality factor $Q_0\simeq 10^8$ of the cavity was used in the calculation. This value corresponds to a surface resistance $R_s \simeq 200 \,\mrm{nOhm}$ of the resonator, which is a rather conservative value, since,  with standard surface preparation techniques, residual resistance values $R_{res} \simeq 1\,\mrm{nOhm}$ have been obtained. Radiation losses will also contribute to the loaded quality factor. In order to minimize them an rf choke was designed (see fig. \ref{fig:dimensioni}). The choke consists in a short--circuited half--wavelength transmission line, tuned to the cavity operating frequency. The first quarter wavelength is a radial waveguide while the second quarter wavelength is a coaxial transmission line that is short circuited at the bottom. 

Other relevant parameters of the system are listed in table \ref{tab:params}.

The mass of the secondary oscillator $m_2$, the rf source power $P_{in}$, and frequency $\om$, and the cavity coupling coefficient $\beta$ were used as freely variable parameters in order to find out the configuration of the transducer that minimizes $h_{min}$ and maximizes the bandwidth.

The result of the optimization are plotted in fig. \ref{fig:larga}: $h_{min}\simeq 2.2\times 10^{-22}\,\mrm{Hz}^{-1/2}$ and $\Delta f\simeq 115\,\mrm{Hz}$ were obtained. These result were calculated with secondary oscillator mass $m_2\simeq 11.6\,\mrm{kg}$, rf source power $P_{in}\simeq 1\,\mu\mrm{W}$, rf source frequency $\om_f=\om_0$, cavity coupling coefficient $\beta\simeq 1$. With these figures the energy stored in the re--entrant cavity is $U\simeq 3\,n\mrm{J}$ and the dissipated rf power is $P_d \simeq 1\,\mu\mrm{W}$. The amplitude of the electric field in the gap is $E_{max}\simeq 8\,\mrm{MV/m}$. This value is well below the limit of vacuum electrical breakdown in rf cavities, which, according to the Kilpatrick criterion \cite{criterion}, is in the range $E_b > 100 \,\mrm{MV/m}$ at the operating frequency of 5 GHz.
Furthermore the peak magnetic field in the cavity is $B_{max}\simeq 0.7\,\mrm{mT}$, which is by far lower than the critical field of niobium at the operating temperature.

\section{Conclusions}
The parametric conversion shifts the detection problem to a very high frequency range, where many noise sources can be made negligible.
The implementation of high frequency electronics has the advantage of a reliable and well developed technology, which exhibits near-SQL behavior in available commercial components. 
Although we understand that the fundamental, physical limits of a parametric transducer are comparable with those of an ordinary capacitive transducer, the parametric one can have technological advantages.
The expected performance shows that the parametric transducer can be competitive with the most advanced SQUID readouts, and can still be a resource for existing and future resonant detectors. 

\begin{table}
\begin{center}
\begin{tabular}{|ccl|}
\hline
$m_1$ & 1135 kg & effective mass of the antenna\\
$L$ & 3 m & length of the antenna\\
$\om_1/(2\pi)$ & 895 Hz & freq. of the primary osc.\\
$\om_2/(2\pi)$ & 900 Hz & freq.of the secondary osc.\\
$Q_1$ & $4\times 10^6$ & quality factor of the bar\\
$Q_2$ & $1.5\times 10^6$ & quality factor of the sec. osc.\\
$\om_0/(2\pi)$ & 5.5 GHz & resonant freq. of the rf cavity\\
$Q_0$ & $1.0\times 10^8$ & unloaded quality factor of the rf cavity \\
$T$ & 0.1 K & thermodynamic temperature\\
$T_e$ & 2 $\hbar\om_0/k_B$ & equivalent temperature of the rf amplifier\\
$S_{osc}$ &  $10^{-12}P_{in}$ & amplitude and phase noise of the rf source\\
$S_{if}$ & 1 $nV/ \sqrt{Hz}$ & input voltage noise of the low frequency amplifier\\
\hline
\end{tabular}
\caption[System parameters]{System parameters}
\label{tab:params}
\end{center}
\end{table}

\end{document}